\documentclass[a4paper,fleqn,usenatbib]{mnras}
\usepackage{amssymb,amsmath,graphicx,psfrag,mathtools}
\usepackage[T1]{fontenc} 
\usepackage{ae,aecompl} 
\usepackage{url}
\title[FRB as Pulsar Lightning]{Fast Radio Bursts as Pulsar Lightning}
\author[J. I. Katz]{
J. I. Katz,$^{1}$\thanks{E-mail: katz@wuphys.wustl.edu} 
\\
$^{1}$Department of Physics and McDonnell Center for the Space Sciences,
Washington University, St. Louis, Mo. 63130 USA 
}
\date{Accepted XXX.  Received YYY; in original form ZZZ} 
\pubyear{2017} 
\date{\today}
\begin{document} 
\label{firstpage} 
\pagerange{\pageref{firstpage}--\pageref{lastpage}} 
\maketitle 
\begin{abstract}
There are striking phenomenological similarities between Fast Radio Bursts
and lightning in the Earth's and planetary atmospheres.  Both have very low
duty factors, $\lesssim 10^{-8}$--$10^{-5}$ for FRB and (very roughly) $\sim
10^{-4}$ for the main return strokes in an active thundercloud.  Lightning
occurs in an electrified insulating atmosphere when a conducting path is
created by and permits current flow.  FRB may occur in neutron star
magnetospheres whose plasma is believed to be divided by vacuum gaps.
Vacuum is a perfect insulator unless electric fields are sufficient for
electron-positron pair production by curvature radiation, a high-energy
analogue of electrostatic breakdown in an insulating gas.  FRB may be
``electrars'' powered by the release of stored electrostatic energy,
counterparts to Soft Gamma Repeaters powered by the release of stored
magnetostatic energy (magnetars).  This frees pulsar FRB models from the
constraint that their power not exceed the instantaneous spin-down power.
Energetic constraints imply that the sources of more energetic FRB have
shorter spin-down lifetimes, perhaps shorter than the three years over which
FRB 121102 has been observed to repeat.
\end{abstract}
\section{Introduction}
The mechanism and astronomical sites of FRB emission remain mysterious,
even after the discovery of the repeating FRB 121102 and its identification
with a dwarf galaxy at redshift $z=0.19$ \citep{C17}.  The requirement that
energies of as much as $10^{40}$ ergs, assuming isotropic emission
(\citet{T13}; see, however, \citet{K17}) be radiated in $\lesssim 1$ ms
points to sources in regions of high power and high energy density 
\citep{K16a}.  The natural candidates are neutron stars, and possible energy
sources are magnetostatic, as in Soft Gamma Repeaters (SGR), or rotational,
as in radio pulsars.

SGR have low duty factors and short rise times analogous to those of FRB,
but their candidacy suffers from the fact that the radiation of SGR is
almost entirely thermal soft gamma rays.  Although at least one SGR has been
observed as a radio source \citep{FKB99}, its emission lasted for weeks,
resembled that of an expanding cloud rather than a fast transient, and
offers no explanation of the FRB phenomenon.  The environment of SGR has an
intense flux of thermal radiation (up to $10^{34}$ erg/cm$^2$/s at the 
neutron star, corresponding to a black body temperature of $3 \times
10^{9\,\circ}$K or 300 keV) which is inhospitable to coherent (or any
nonthermal) radio radiation because of rapid Comptonization of energetic
electrons.  In addition, one FRB was fortuitously the subject of radio
observation \citep{TKP16} during its giant outburst, but no FRB was
observed.  As a Galactic object, had it been the source of a FRB an
extraordinarily strong signal, even in the far sidelobes of a radio
telescope, would have been expected
\citep[for possible loopholes see][]{K16b}.

Giant pulses from radio pulsars may be more plausible explanations of FRB.
The fact that nanoshots from two Galactic radio pulsars \citep{S04,HE07} had
brightness temperatures exceeding even those of FRB points in this
direction.  No detailed mechanism has been proposed, but after nearly 50
years in which theorists have failed to present a convincing detailed
mechanism of even ``ordinary'' coherent radio pulsar emission, this should
not be a fatal objection.  Even the basic electrodynamics of pulsar
magnetospheres remains controversial \citep{M90,M03,MY16}.

Pulsar magnetospheres contain ``gaps'' in which ${\vec E}\cdot{\vec B} \ne
0$ \citep{MY16}.  In these gaps intense electric fields accelerate electrons
and positrons.  Their curvature radiation consists of gamma-rays of
sufficiently high energy to produce further electron-positron pairs by a
single-photon process in the strong neutron star magnetic field
\citep{RS75}.  The purpose of this paper is to point out an analogy between
the breakdown of a resistive atmosphere in a lightning stroke and the
breakdown of the resistive vacuum in a gap in a pulsar magnetosphere.

This is only an analogy.  The fundamental physical processes are different
and the parameter regimes differ by many orders of magnitude.  Estimates are
necessarily qualitative, or, at best, order-of-magnitude.  Even were it
possible to construct a more quantitative theory, the range of plausible
pulsar parameters is several orders of magnitude in both spin frequency and
magnetic field.  It includes such hypothetical objects as neutron stars with
both surface fields of $\sim 10^{15}$ gauss and millisecond periods.  The
constraint of a lifetime $> 3\,$y inferred from the repeating FRB 121102
need not apply to non-repeating FRB that might spin down much faster.  

\citet{IS91a,IS91b} have suggested ``lightning'' triggered by external
$\gamma$-rays as a source of RRAT emission.  An important difference between
their work and this paper is the suggestion here that ``lightning'' is
powered by stored electrostatic energy, and thus may involve and radiate a
power greater, perhaps by orders of magnitude, than the spindown power.
This is likely necessary to explain FRB, but not to explain RRAT.

Although there is a fairly detailed semiquantitative understanding of 
terrestrial cloud charging and lightning, it is unlikely that we would have
predicted their existence had we not known of it.  This should remind us
that pulsar magnetospheres may also behave in unexpected ways.  Unable to
predict their phenomenology, our understanding may be limited to delineating
the range of possibilities permitted by physical laws. 
\section{Lightning} 
Lightning has been the subject of scientific study since the work of Ben
Franklin nearly 300 years ago.  The historic and modern literatures are
reviewed by \citet{U69} and \citet{RU03}.  Cumulonimbus clouds
(thunderheads) are charged by differential motion among oppositely charged
water droplets, graupel (partially frozen hail) and ice.  Potential drops
are $\sim 10^8\,$V over 5--10 km and discharges carry $\sim 10$ Coulombs,
releasing $\sim 10^9\,$J of electrostatic energy.  This is $\gtrsim 10\%$ of
the total electrostatic energy of the cloud, a capacitor of dimension $\sim
10\,$km in each of three dimensions (capacitance $\sim 1\,\mu$F).  Not only
does the air break down along the visible channel of the lightning stroke
(bolt), but there must be sufficient conductivity outside the channel itself
to collect a substantial fraction of the charge distributed throughout the
cloud.
\section{Pulsar Magnetospheres}
Much less is known about pulsar magnetospheres.  Unlike thunderclouds, they
cannot be imaged or probed; we only observe their radiation that is directed
toward us and its variation as the neutron star rotates.  Both their
structure and the mechanisms of production of pulsar radiation remain
controversial, nearly 50 years after their discovery \citep{M90,M03,MY16}.

It is generally assumed that the magnetic fields of pulsar magnetospheres
are determined by currents within the dense matter of the neutron star that
change only on a very long ($> 10^6\,$y) resistive time scale.  This is in
contrast to SGR, that are believed to result from the sudden release of
magnetostatic energy in a reorganization of the magnetic field
\citep{K82,TD92,TD95} resulting from reconnection of magnetospheric
currents.  These currents may be relics of the formation of the neutron star
that dissipate in $\lesssim 10^4\,$y, the empirical lifetimes of SGR
\citep{K16b}. 
\subsection{States}
If there are two or more possible states, with different charge
distributions and electrostatic energies, of a pulsar magnetosphere, with
the more energetic state metastable, then there may be sudden transitions
between them with the release of their energy difference.  This would be
analogous to the transition of a thundercloud from a more to a less
energetic charge distribution as charge flows in a lightning bolt.  

Pulsar magnetospheres are highly relativistic.  The characteristic plasma
density of a co-rotating magnetosphere with angular velocity $\omega_m$
\citep{D47,D55,HB65,GJ69} is
\begin{equation}
\label{GJ}
n_e = {{\vec B} \cdot {\vec \omega_m} \over 2 \pi ec} \approx 1.1 \times
10^{17} {\vec B}_{15} \cdot {\vec \omega}_{m,4}\ \text{cm}^{-3},
\end{equation}
where $B_{15} \equiv B/10^{15}\,\text{gauss}$ and $\omega_{m,4} \equiv
\omega_m/ 10^4\,\text{s}^{-1}$.  The ratio of its magnetic energy density to
its characteristic rest mass energy density
\begin{equation}
\label{Gamma}
\Gamma \equiv {B^2/8\pi \over n_e m_e c^2} = {B e \over 4 \omega m_e c}
\approx 4.4 \times 10^{17} {B_{15} \over \omega_4} \ggg 1.
\end{equation}
This is very large for all pulsars, even low field millisecond pulsars.

Because magnetospheres are relativistic, the release of energy in a
transition between two states occurs in a time $\sim R/c \sim 3 \times
10^{-5}\,$s, where $R$ is the neutron star's radius, consistent with
empirical upper bounds to FRB widths at their sources (before propagation
broadening).  There is no generally accepted solution for pulsar
magnetospheres (much less two solutions, one of which is more energetic but
metastable) so we only make order of magnitude estimates.

The fundamental equation of a co-rotating force-free magnetosphere is
\begin{equation}
\label{ecorot}
{\vec E} + {{\vec \omega_m} \times {\vec r} \over c} \times {\vec B} = 0,
\end{equation}
where the co-rotation frequency $\omega_m$ need not be the same as the
neutron star's rotation frequency $\omega$.  \citet{RS75} present an
explicit solution for an aligned rotator with $\omega_m \ne \omega$.  Within
a connected corotating region ${\vec E} \perp {\vec B}$; $E_\parallel = 0$.
Distinct corotating regions with differing ${\vec \omega}_m$ are separated
by gaps in which $E_\parallel \ne 0$.

From Eq.~\ref{ecorot} the electrostatic energy may be estimated
\begin{equation}
\label{encorot}
{\cal E}_{el} \sim {\omega_m^2 B^2 R^5 \over 2 c^2} \sim 5 \times 10^{46}
\omega_m^2 B_{15}^2\ \text{ergs},
\end{equation}
and the energy released in a transition between an initial $\omega_{m,i}$
and a final $\omega_{m,f}$
\begin{equation}
\label{deltae}
\Delta {\cal E}_{el} \sim (\omega_{m,i}^2 - \omega_{m,f}^2) {B^2 R^5 \over
2 c^2} \sim 5 \times 10^{46} \Delta \omega_4^2 B_{15}^2\ \text{ergs},
\end{equation}
where the fractional change in $\omega_m^2$ $\Delta \equiv (\omega_{m,i}^2 -
\omega_{m,f}^2) / \omega^2$.  $\Delta$ is unknown, as is the efficiency
$\epsilon$ of conversion of this energy into coherent FRB emission.  Both
may be $\ll 1$, but if both $\omega$ and $B$ are large there may still be
sufficient energy to power FRB.
\subsection{Capacitor}
Alternatively, treat the magnetosphere as a capacitor with capacitance $C
\sim R^2/h$, where $h$ is the width of its gap, charged to the
characteristic voltage in a polar cap model \citep{RS75} $V \sim \omega^2
R^3 B/c^2$.  The resulting energy would be
\begin{equation}
\label{capacitor}
{\cal E}_{cap} \sim {\omega^4 B^2 R^8 \over 2 h c^4} \sim 5 \times
10^{45} {R \over h} B_{15}^2 \omega_4^4\ \text{ergs},
\end{equation}
The value of $h$ is unknown, though it cannot exceed $R$.  For sufficiently
strongly magnetized and rapidly rotating neutron stars the electrostatic
energy may be more than sufficient to power FRB, the most energetic of which
radiated, assuming isotropic emission, $10^{40}$ ergs, even if the
efficiency $\epsilon \ll 1$.
\subsection{Parameters}
The required values of $B$ and $\omega$ are found in known pulsars, but not
in combination.  That is attributed to their ages (Eq.~\ref{tau}, below).
A pulsar-FRB might be as young as the time required for a supernova remnant
to dissipate to transparency, and therefore could have both large $B$ and
large $\omega$.  Neutron star formation by collapse without expulsion of a
remnant or a visible supernova may also be possible (codes have had more
difficulty producing a supernova than neutron star formation without one).
If so, the only lower bound on the duration of FRB activity would be the
observed activity of a repeating FRB, permitting simultaneously large $B$
and large $\omega$.  The duration could be even shorter, and $B$ and
$\omega$ even larger, for non-repeating FRB, for which no lower bound on
their duration of activity can be established.
\subsection{Vacuum Breakdown}
In the absence of a magnetic field, very large electric fields are not
sustainable.  For example, Eq.~\ref{capacitor} assumes an electric field
(in cgs units, statvolts/cm)
\begin{equation}
E \sim {\omega^2 R^3 B \over hc^2} \sim 10^{14} {R \over h} \omega_4^2
B_{15}.
\end{equation}
If $\omega_4^2 B_{15} > 0.02 h/R$ this exceeds the 
Sauter-Heisenberg-Euler-Schwinger \citep{S31,HE36,S51} vacuum pair breakdown
field (in statvolts/cm) 
\begin{equation}
\label{Schwinger}
E_{SHES} \approx 0.05 {m_e^2 c^3 \over e \hbar} \approx 2 \times
10^{12}.
\end{equation}

Pair production cascades driven by curvature radiation may break down the
insulating vacuum at lower (but still very high) fields \citep{RS75}.
``Schwinger sparks'', produced by fields exceeding the limit of
Eq.~\ref{Schwinger}, have been proposed \citep{SY15} as the source of pulsar
nanoshots.  The FRB model proposed here does not require that limit to be
exceeded.

If the available electrostatic energy in a capacitor model is limited by the
energy density $E_{SHES}^2/8\pi$, then
\begin{equation}
\label{SHES}
{\cal E}_{cap} \sim {1 \over 2} h R^2 E_{SHES}^2 \sim 2 \times
10^{42} {h \over R}\ \text{ergs}.
\end{equation}
This is less than the limit of Eq.~\ref{capacitor} for rapidly spinning or
strongly magnetized neutron stars, particularly if $h$ is small.  It is
still sufficient to power the observed FRB provided $h \sim 10^4$ cm
(similar to gap widths estimated by \citet{RS75}), the electrostatic energy
difference between initial and final configurations is comparable to that
given by Eq.~\ref{SHES}, and emission is efficient and roughly isotropic.
Collimated emission \citep{K17} would mitigate the energy requirement.  The
efficiency is, of course, unknown.

There can be no pair production unless $E_\parallel \ne 0$.  If $E_\parallel
\ne 0$ in one frame, the Lorentz invariance of ${\vec E} \cdot {\vec B}$
implies that $E_\parallel \ne 0$ in all frames.  In gaps, where there is a
parallel component of $\vec E$, breakdown can occur and $E$ has a limit like
Eq.~\ref{Schwinger} \citep{KP06}.

If there be a magnetic field with $B^2 > E^2$ and ${\vec E} \perp {\vec B}$,
as expected in a neutron star magnetosphere outside of ``gaps'', the Lorentz
invariance of $B^2 - E^2$ and of ${\vec E} \cdot {\vec B} = 0$ implies the
existence of a frame in which ${\vec E} = 0$, and there can be no pair
production.  Pair production cannot set an upper bound on $E$, other than
$B$, in corotating regions in which ${\vec E} \perp {\vec B}$.  The limit
of Eq.~\ref{SHES} is not applicable to the electrostatic energy stored, and
released, in regions where ${\vec E} \perp {\vec B}$.  Energies as great as
those of Eq.~\ref{deltae} may be available to power FRB.
\section{Plasma Frequency}
FRB might be produced by conversion of longitudinal electrostatic plasma
waves, propagating along the magnetic field (with the same dispersion
relation as in a nonmagnetic plasma), to transverse electromagnetic waves.
The plasma frequency corresponding to the characteristic plasma density
Eq.~\ref{GJ}, allowing for the possibility that the electrons have
relativistic speeds, reducing their acceleration by electric fields parallel
to their velocities (that are parallel to $\vec B$) by a factor
$\gamma^{-3}$, is 
\begin{equation}
\label{plasma}
\begin{split}
\nu_p &= {1 \over 2\pi} \sqrt{{4 \pi n_e e^2 \over m_e}\left\langle{1 \over
\gamma^3}\right\rangle} = {1 \over 2 \pi} \sqrt{{2 e {\vec B}\cdot{\vec
\omega_m} \over m_e c}\left\langle{1 \over \gamma^3}\right\rangle}\\
&\approx 3.0 \times 10^{12} \sqrt{{\vec B}_{15} \cdot {\vec \omega}_{m,4}
\left\langle{1 \over \gamma^3}\right\rangle}\ \text{Hz},
\end{split}
\end{equation}
where $\gamma$ is a mean Lorentz factor of electron or positron motion along
the magnetic field, assumed to be the direction of propagation of the
longitudinal plasma wave.
\section{Spin-Down Time}
The spin-down time of a rotating magnetic neutron star, taking the magnetic
moment as perpendicular to its angular momentum, is
\begin{equation}
\label{tau}
\tau = {3 c^3 I \over 2 B^2 \omega^2 R^6} \approx {4 \times 10^2\,\text{s}
\over B_{15}^2 \omega_4^2},
\end{equation}
where $I \approx 10^{45}\,$g-cm$^2$ is its moment of inertia.  The product
$B \omega$ appears as in Eq.~\ref{plasma} for the plasma
frequency, so that $\tau$ can be expressed as a function of $\nu_p$:
\begin{equation}
\label{spindown}
\tau = {3 \over 2} {c I e^2 \over (2 \pi \nu_p)^4 m_e^2 \gamma^6} \approx
8 \times 10^{15} \left({\nu_p \over \text{1\,GHz}}\right)^{-4} \gamma^{-6}\
\text{s}.
\end{equation}

If the plasma frequency is that of nonrelativistic electrons, then the
spin-down time is comparable to that of a typical observed Galactic pulsar.
Such pulsars are numerous, with $\sim 10^4$ in the Galaxy, while FRB are
extremely rare, with a detected event rate $\sim 10^{-5}$/galaxy-y, arguing
against the hypothesis that FRB are produced by pulsars resembling observed
Galactic pulsars.  If the plasma is relativistic, the sensitivity of $\tau$
to $\gamma$ is so extreme that no quantitative conclusion can be drawn from
Eq.~\ref{spindown}.

Eq.~\ref{tau} can be compared to the observed lower bound on the lifetime of
the repeating FRB 121102 to infer $\omega_4^2 B_{15}^2 < 4 \times 10^{-6}$.
Comparing to Eq.~\ref{deltae} then leads to an upper bound on the available
energy $\sim 10^{41}$ ergs, consistent with the observed energy of its FRB
of $\sim 10^{38}$ ergs.  Equivalently, the spin-down time of a pulsar
emitting FRB of energy ${\cal E}_{FRB}$ with efficiency $\epsilon$ is
\begin{equation}
\label{life}
\tau \approx 2 \times 10^9 {\Delta \epsilon \over {\cal E}_{FRB40}}\
\text{s}
\end{equation}
where ${\cal E}_{FRB40} \equiv {\cal E}_{FRB}/10^{40}$ ergs.  Unfortunately,
we cannot estimate $\Delta$ or $\epsilon$, and if FRB emission is collimated
\citep{K17} ${\cal E}_{FRB}$ may be much less than that inferred with the
assumption of isotropic emission.  Therefore, it is not possible to make
quantitative predictions from Eq.~\ref{life}.
\section{Discussion}
This paper proposes that FRB are ``electrars'' powered by electrostatic
energy, in analogy to SGR that are powered by magnetostatic energy
(``magnetars'').  In each case energy is released by a transition between
two configurations, the more energetic of which is metastable.  Neither of
these phenomena were predicted before their discovery; Nature is cleverer
than scientists at finding ways to tap stored energy.

The energetic requirements may be met in the magnetospheres of fast
high-field pulsars.  None of these are known, but their spin-down lifetimes
would be very short, conceivably as short as minutes, and most may be hidden
within opaque young supernova remnants.  This is consistent with the extreme
rarity of FRB.

A weakness of the hypothesis suggested here is that there is no theory that
demonstrates the existence of two (or more) magnetospheric configurations
with distinct energies, nor that such an energy difference can be released
suddenly.  Without an understanding of the magnetospheric structure of
pulsars that may not be a fatal flaw, but it is a serious one.  There is,
however, a precedent from SGR: Although any magnetosphere has a lower energy
state in which there is no magnetic field, {\it a priori\/} the only path to
it is by very slow resistive dissipation that does not make outbursts.  The
fact that sudden magnetic reconnection can and does occur is empirical and
not based on fundamental understanding.  Perhaps electrostatic energy too
may be released in a burst, as it is in a thundercloud.

Combination of Eq.~\ref{deltae} with Eq.~\ref{tau} implies that the more
energetic FRB are likely to have lifetimes much shorter than the lower bound
of 3 years observed for the repeating FRB 121102.  This conclusion cannot be
quantified without some knowledge of $\Delta$ and $\epsilon$, but it is
qualitatively consistent with the absence of repetitions of more energetic
FRB.  The bursts of FRB 121102 had particularly low energy; the mean fluence
of 17 bursts \citep{S16,Sc16} was 0.275 Jy-ms.  Assuming isotropic emission
at $z = 0.19$, the mean energy radiated, integrated over the 1214--1537 MHz
band of observation, was $1.0 \times 10^{38}\,$ergs, less than the inferred
energies of $\sim 3 \times 10^{38}\text{--}10^{40}\,$ergs of other FRB.

The efficiency $\epsilon$ of coherent radio emission of pulsars is low.
However, when the bunching of charges is extreme, as it must be to explain
FRB, radiation is rapid.  For example, \citet{KLB17} estimate energy loss
times $\sim 10^{-15}\,$s.  Then, in order for the emission to be at GHz
frequencies (rather than $\sim 10^{15}$/s) radiative losses must be very
closely balanced by the accelerating electric field on the energy loss time
scale.  This explains the efficient radiation as a self-consistent property
of extreme charge bunching.

If more energetic FRB are produced by pulsars with particularly short
spin-down times, searches for repetitions must be conducted immediately
after their first occurrence.  The observability of FRB may depend on the
rapid dissipation of a supernova remnant, a patchy remnant, or neutron star
formation without a supernova at all.  These requirements may help explain
the fact that the FRB rate, despite the possibility of repetitions, is less
than $10^{-3}$ of the supernova rate.

The broad range (approximately -10 to +14) of spectral indices observed
\citep{S16} in bursts from the repeating FRB 121102 might appear to indicate
comparatively narrow-band emission with power peaking at frequencies
sometimes above, sometimes below, and sometimes in the middle of the band
of observation.  
%
However, the rapid variation of burst widths indicates that scintillation is
strong.  \citet{S16} report widths of $6.6 \pm 0.1\,$ms and $3.06 \pm
0.04\,$ms for two bursts (their bursts 8 and 11) 413 s apart.  A factor of
two variation in broadening suggests ${\cal O}(1)$ variation in fluence and
frequencies of spectral peaks and minima, indicating that the variation in
fitted spectral indices is the result of scintillation, not an intrinsic
property of the source.

\bsp 
\label{lastpage} 
\end{document}